\documentclass[preprintnumbers,floatfix,twocolumn,nofootinbib]{revtex4}

\usepackage[tbtags]{amsmath}  
\usepackage{amssymb}          
\usepackage{bm}               
\usepackage{graphicx}         
\usepackage{hhline,multirow}  
\usepackage{dcolumn}          

\newcommand{\Pslash}{P \hspace{-0.24cm} / \,}
\newcommand{\kslash}{k \hspace{-0.21cm} / \,}
\newcommand{\nslash}{n \hspace{-0.24cm} / \,}
\newcommand{\vslash}{v \hspace{-0.21cm} / \,}
\newcommand{\Sslash}{S \hspace{-0.22cm} / \,}
\newcommand{\Dslash}{D \hspace{-0.22cm} / \,}
\newcommand{\epsslash}{\varepsilon \hspace{-0.18cm} / \,}

\allowdisplaybreaks[2]

\begin{document}


\preprint{JLAB-THY-09-1033}

\title{What can break the Wandzura--Wilczek relation?}

\author{Alberto~Accardi$^{a,b}$, Alessandro~Bacchetta$^{a,c}$,
	W.~Melnitchouk$^{a}$, Marc~Schlegel$^{a}$} 
\affiliation{
$^a$Jefferson Lab, Newport News, VA 23606, USA \\
$^b$Hampton University, Hampton, VA 23668, USA \\
$^c$Universit\`a degli Studi di Pavia, 27100 Pavia, Italy
}

\begin{abstract}
We analyze the breaking of the Wandzura--Wilczek relation for the 
$g_2$ structure function, emphasizing its connection with transverse 
momentum dependent parton distribution functions.  We find that the 
relation is broken by two distinct twist-3 terms, and clarify how
these can be separated in measurements of double-spin asymmetries in
semi-inclusive deep inelastic scattering.  Through a quantitative
analysis of available $g_2$ data we also show that the breaking of 
the Wandzura--Wilczek relation can be as large as 15--40\% of the
size of $g_2$.
\end{abstract}



\maketitle

\section{Introduction}

The spin structure of the nucleon remains one of the most challenging
and controversial problems in hadronic physics
\cite{Bass:2004xa,Kuhn:2008sy,Burkardt:2008jw}.  Experimentally it is
now known, through many careful measurements of the nucleon's $g_1$
structure function, that quarks carry only some 30\% of the proton's
longitudinal spin, a feature which is now qualitatively understood
\cite{Myhrer:2007cf}.  Moreover, polarized $pp$ scattering observables
\cite{Morreale:2009zr} and open charm production in deep inelastic
scattering \cite{Alekseev:2009ey} suggest that gluons carry an even
smaller fraction of the longitudinal spin.  Presumably, the remainder
arises from quark and gluon orbital angular momentum.

Although less attention has been paid to it, there are a number of
intriguing questions associated with the transverse spin structure
of the nucleon.  An example is the study of the $g_2$ structure 
function, which only in recent years has been probed experimentally 
with high precision.  Unlike all other inclusive deep-inelastic 
scattering (DIS) observables, the $g_2$ structure function is unique
in directly revealing information on the long-range quark-gluon 
correlations in the nucleon.   In the language of the operator 
product expansion (OPE) these are parametrized through matrix elements 
of higher twist operators, which characterize the strength of 
nonperturbative multi-parton interactions.  (In the OPE ``twist''
is defined as the mass dimension minus the spin of a local operator.)
In other inclusive structure functions higher twist contributions are
suppressed by powers of the four-momentum transfer squared $Q^2$,
whereas in $g_2$ these appear at the same order as the leading twist.

As discussed by Wandzura and Wilczek \cite{Wandzura:1977qf}, the
leading twist contribution to the $g_2$ structure function, which is
denoted by $g_2^{\rm WW}$, can be expressed in terms of the leading 
twist (LT) part of the $g_1$ structure function,
%
\begin{equation}
g_2^{\rm WW}(x_B)
= - g_1^{\rm LT}(x_B) + \int_{x_B}^1 \frac{dy}{y} g_1^{\rm LT}(y)\ ,
\label{eq:g2WWdef}
\end{equation}
where $x_B$ is the Bjorken scaling variable, and we suppress the
explicit dependence of the structure functions on $Q^2$.
The Wandzura--Wilczek (WW) relation asserts that the total $g_2$
structure function is given by the leading twist approximation
(\ref{eq:g2WWdef}),
\begin{equation}
g_2(x_B) \stackrel{?}{=} g_2^{\rm WW}(x_B)\ ,
\label{eq:WWrel}
\end{equation}
which would be valid in the absence of higher twist contributions.
In this case the $g_2$ structure function would also 
satisfy the Burkhardt--Cottingham (BC) sum rule
\cite{Burkhardt:1970ti},
\begin{equation}
\int_0^1 dx_B\ g_2(x_B) = 0\ .
\label{eq:BC}
\end{equation}
Its violation would also signal the presence of twist-3 or higher contributions.
Unlike the WW relation, however, 
the validity of the BC sum rule (which is yet to be
conclusively demonstrated experimentally 
\cite{Anthony:2002hy,Amarian:2003jy}) would not necessarily imply that
higher twist terms vanish \cite{Jaffe:1989xx,Jaffe:1990qh}.

In this paper we explore the physics that can lead to the breaking of
the WW relation in QCD, preliminary results for which have appeared
in Ref.~\cite{Accardi:2009nv}.  In Sec.~II we present a detailed 
theoretical analysis of quark-quark and quark-gluon-quark correlation
functions, and discuss the so-called Lorentz invariance relations and 
equations of motion relations.  From these we show that the WW relation
is valid if pure twist-3 and quark mass terms are neglected, in 
agreement with OPE results.  We find that there
are {\it two} distinct contributions with twist 3, denoted by 
$\widetilde g_T$ and $\widehat g_T$, which correspond to two 
different ``projections'' of the quark-gluon-quark correlator.
An explicit demonstration of our findings is made for the case of
a point-like quark target, which shows that the twist-3 terms
can in principle be as large as the twist-2 terms.

In Sec.~III we discuss the phenomenology of the WW relation for both
the proton and neutron, and find that the available data from SLAC
and Jefferson Lab indicate a breaking of the WW relation at the level
of 15--40\% of the size of $g_2$ within the 1-$\sigma$ confidence level.
The two twist-3 terms can be separated by measuring, in addition to 
$g_2$, the function $g_{1T}^{(1)}$ in semi-inclusive DIS, as we outline 
in Sec.~IV.  There we explain the importance of measuring the two 
twist-3 functions $\widetilde g_T$ and $\widehat g_T$ separately, and 
the insight which this can bring, for example, to understanding the 
physics of quark-gluon-quark correlations~\cite{Burkardt:2008ps},
or to determining the QCD evolution kernel for $g_2$ and the large
momentum tails of transverse momentum distributions (TMDs).

Finally, in Sec.~V we briefly summarize our findings.
Some technical details for the analysis with a non-lightlike
Wilson line and the model calculation of parton correlation
functions are presented in the appendices.

\section{Theoretical analysis} 

In this section we set forth the framework for our analysis of the
WW relation by first defining quark-quark correlation functions and
examining their most general Lorentz and Dirac decomposition.
This is followed by a discussion of quark-gluon-quark correlators, 
and of the Lorentz invariance and equations
of motion relations from which a generalization of the WW relation is derived.

\subsection{Parton correlation functions}

The quark-quark correlator for a quark of momentum $k$ in a nucleon
with momentum $P$ and spin $S$ is defined as
\begin{equation} 
\begin{split} \label{e:corr1}
&\Phi^a_{ij}(k,P,S; v)  =
 \int \frac{d^4 \xi}{(2 \pi)^4} \, e^{i k \cdot \xi} \,
\\
& \quad
\times
 \langle P,S \, | \, \overline{\psi}^{\, a}_j(0) \,
 {\cal W}^{v}_{(0,\infty)} \, 
 {\cal W}^{v}_{(\infty,\xi)} \, 
\psi^a_i(\xi) \, | \, P,S \rangle\, ,
\end{split} 
\end{equation}
where the quark fields $\psi^a_i$ are labeled by the flavor index $a$
and Dirac index $i$.  For ease of notation, the Dirac and flavor
indices will be suppressed in the following.  The operator
${\cal W}^v_{(0,\infty)}$ represents a Wilson line (or gauge link)
from the origin to infinity along the direction specified by the vector
$v$, and is necessary to ensure gauge invariance of the correlator.
The gauge links contain transverse pieces at infinity
\cite{Belitsky:2002sm,Boer:2003cm} and their precise form depends
on the process~\cite{Collins:2002kn,Bomhof:2006dp}. 
In a covariant gauge, the dependence of the correlator $\Phi$ on $v$
is evident from the presence of the Wilson line in the direction
conjugate to $v$. 
In light-cone gauges the vector $v$ is orthogonal to the gauge field
$A$, $v\cdot A=0$, and the dependence on $v$ 
appears explicitly only in the gauge field propagators.

In tree-level analyses of semi-inclusive DIS (SIDIS)
\cite{Mulders:1995dh,Bacchetta:2006tn} or the Drell-Yan process
\cite{Tangerman:1994eh,Boer:1999mm,Arnold:2008kf} $v$ is identified
with the light-cone vector $n_-$, where $n_-^2 = 0 = n_+^2$ and
$n_-\cdot n_+ = 1$, with $n_+$ the corresponding orthogonal
light-cone vector proportional to $P$ (up to mass corrections).
However, factorization theorems beyond tree-level
\cite{Collins:1981uk,Ji:2004wu,Collins:2004nx,Collins:2007ph} demand
a slightly non-lightlike vector $v$ in order to regularize light-cone
divergences.
We leave a more detailed discussion of the effect of the choice of $v$
to Appendix~\ref{a:v} and consider $v=n_-$ unless otherwise specified.

The correlator $\Phi$ can be parametrized in terms of structures built
from the four vectors $P$, $S$, $k$ and $v$. Its full decomposition
has been studied in Ref.~\cite{Goeke:2005hb} (and further generalized in
Ref.~\cite{Meissner:2009ww}).  It contains 12 scalar functions $A_i$
already known from Refs.~\cite{Mulders:1995dh,Tangerman:1994bb}, and
20 scalar functions $B_i$ which are multiplied by factors depending
explicitly on $v$, which were first introduced in
Ref.~\cite{Goeke:2003az} and called parton correlation functions (PCFs)
in Ref.~\cite{Collins:2007ph}.  For brevity we consider only those
terms of the full decomposition \cite{Goeke:2005hb} which are necessary
for the present analysis,
\begin{equation} 
\begin{split} 
 \Phi&(k,P,S; v) = M\Sslash \gamma_5 A_6
		  + \frac{k\cdot S}{M} \Pslash  \gamma_5 A_7 
\\ &
+\frac{k\cdot S}{M} \kslash \gamma_5 A_8
+ M \frac{(S \cdot v)}{(P \cdot v)} \Pslash \gamma_5 B_{11}
+ M \frac{(S \cdot v)}{(P \cdot v)} \kslash \gamma_5 B_{12}
\\  &
+ M \frac{(k \cdot S)}{(P \cdot v)} \vslash \gamma_5 B_{13}
+ M^3\frac{(S \cdot v)}{(P \cdot v)^2} \vslash \gamma_5 B_{14}
+ \cdots\ ,
\label{eq:decomp} 
\end{split} 
\end{equation} 
where the nucleon mass $M$ is explicitly included to ensure that
all PCFs have the same mass dimension.
(Any other hadronic scale, such as $\Lambda_{\rm QCD}$,
can be chosen, but we follow the choice used in the TMD 
literature~\cite{Mulders:1995dh}.)

The PCFs $A_i$ and $B_i$ are in principle functions of the scalar
products $P \cdot k$, $k^2$, $P \cdot v$, $k \cdot v$ and $v^2$.
However, because the correlator $\Phi$ is invariant under the scale
transformation $v \to \lambda v$, where $\lambda$ is a constant,
the PCFs can only depend on ratios of the scalar products,
$P \cdot k$, $k^2$ and $k \cdot v / P \cdot v$.  We therefore choose
the PCFs to depend on the parton virtuality $\tau \equiv k^2$,
on $\sigma \equiv 2 P \cdot k$, and on the parton momentum fraction
$x = k\cdot n_-/P\cdot n_-$. 
We emphasize that the explicit dependence on $x$ is induced in general by the
$v$ dependence of the correlator $\Phi$.

These considerations apply even when the correlator is integrated over
the parton transverse momentum, and in fact the $B_i$ terms give
contributions also to standard collinear parton distribution functions
(PDFs), such as the helicity distribution --- see Eq.~\eqref{e:g1_AB}
below.  However, when the correlator is fully integrated over $d^4k$
the $B_i$ no longer contribute; indeed
\begin{align}
  \int d^4k \, \Phi(k,P,S; v)  =
    \langle P,S \, | \, \overline{\psi}(0)\,
    \psi_i(0) \, | \, P,S \rangle \ ,
\label{e:phi_int}
\end{align}
and the dependence of the integral on $v$ disappears because
${\cal W}^v_{(0,\infty)} {\cal W}^v_{(\infty,0)} = 1$.

\begin{widetext}
In TMD factorization the relevant objects are the integrals 
of $\Phi(k,P,S; v)$ over $k^- = k_\mu n_+^\mu$,
\begin{align} 
\Phi(x,\bm{k}_T) = \int d k^-\,\Phi(k,P,S; v) = 
 \int \frac{d \xi^- d^2\xi_T}{(2 \pi)^3} \, e^{i k \cdot \xi} \,
\langle P,S \, | \, \overline{\psi}(0) \,
 {\cal W}^v_{(0,\infty)} \, 
 {\cal W}^v_{(\infty,\xi)} \, 
\psi(\xi) \, | \, P,S \rangle\Big|_{\xi^+=0}\, .
\end{align} 
It is also useful to define the $\bm{k}_T$-integrated correlators
\begin{align}
\begin{split} 
\Phi(x) &= \int d^2 \bm{k}_T\, \Phi(x,\bm{k}_T)
=  \int \frac{d \xi^-}{2 \pi} \, e^{i k \cdot \xi} \,
\langle P,S \, | \, \overline{\psi}(0) \,
 {\cal W}^v_{(0,\infty)} \, 
 {\cal W}^v_{(\infty,\xi)} \, 
\psi(\xi) \, | \, P,S \rangle\Big|_{\xi^+=\xi_T=0}\, 
\\ &
\stackrel{\text{LC}}{=}
 \int \frac{d \xi^-}{2 \pi} \, e^{i k \cdot \xi} \,
\langle P,S \, | \, \overline{\psi}(0) \,
\psi(\xi) \, | \, P,S \rangle\Big|_{\xi^+=\xi_T=0}\, ,
\end{split}  
\\
\begin{split} 
\Phi_{\partial}^\alpha(x) &= \int d^2\bm{k}_T\,
	k_T^\alpha \Phi(x,\bm{k}_T)
= \int \frac{d \xi^-}{2 \pi} \, e^{i k \cdot \xi} \,
\langle P,S \, | \, \overline{\psi}(0) \,
 {\cal W}^v_{(0,\infty)} \, i \partial_T^\alpha \,
 {\cal W}^v_{(\infty,\xi)} \, 
\psi(\xi) \, | \, P,S \rangle\Big|_{\xi^+=\xi_T=0}\, ,
\\
&\stackrel{\text{LC}}{=}
 \int \frac{d \xi^-}{2 \pi} \, e^{i k \cdot \xi} \,
\langle P,S \, | \, \overline{\psi}(0) \, i \partial_T^\alpha \,
\psi(\xi) \, | \, P,S \rangle\Big|_{\xi^+=\xi_T=0}\, .
\end{split}  
\end{align} 
where LC refers to the correlators in the light-cone gauge.
The correlator $\Phi_{\partial}^\alpha$ actually depends on the
detailed form of the Wilson line, and changes, for example, between
the SIDIS and Drell--Yan processes.  However, for our discussion this
will not be relevant and we can consider the average between the
correlator for SIDIS and Drell--Yan~\cite{Boer:2003cm}.
\end{widetext}

For any correlator, we can introduce the Dirac projections
\begin{equation}
\Phi^{[\Gamma]} \equiv \frac12 \text{Tr} [\Gamma \Phi]\ ,
  \label{Phiproje}
\end{equation} 
where $\Gamma$ is a matrix in Dirac space.
The transverse momentum dependent parton distribution functions then
appear as terms of the general decomposition of the projections
$\Phi^{[\Gamma]}(x,\bm{k}_T)$, the full list of which can be found
in Refs.~\cite{Goeke:2005hb,Bacchetta:2006tn}.  
Usually a TMD is defined to have ``twist'' equal to $n$ if in the
expansion of the correlator it appears at order $(M/P^+)^{n-2}$,
where $P^+ = P_\mu n_-^\mu$.
In physical observables, TMDs of twist $n$ appear with a suppression
factor $(M/Q)^{n-2}$ compared to twist-2 TMDs. 
We finally note that at present TMD factorization for SIDIS has been
proven for twist-2 TMDs only~\cite{Ji:2004wu}, and problems are known
to occur at twist 3, indicating that the formalism may not yet be
complete~\cite{Gamberg:2006ru,Bacchetta:2008xw}.

For the following discussion we shall need the definitions of certain TMDs  
(note that here and in the following $\alpha$ is restricted to be a
transverse index)~\cite{Bacchetta:2006tn} 
\begin{align} 
  \Phi^{[\gamma^+ \gamma_5]}(x,\bm{k}_T)
& = S_L \, g_{1L}(x,\bm{k}_T^2) 
  + \frac{\bm{k}_T \cdot \bm{S}_T}{M}\, g_{1T}(x,\bm{k}_T^2)\ ,
    \label{eq:TMD_g1} 
\\ 
\begin{split} 
  \Phi^{[\gamma^{\alpha}\gamma_5]}(x,\bm{k}_T)
& = \frac{M}{P^+} 
    \bigg[S_T^{\alpha} \, g_T(x,\bm{k}_T^2) 
    + S_L \, \frac{k_T^{\alpha}}{M}\, g_L^{\perp}(x,\bm{k}_T^2)
\\ &\quad \quad
    - \frac{k_T^{\alpha}\, k_T^\rho
      +\frac{1}{2}\,\bm{k}_T^2\,g^{\alpha\rho}_T}{M^2} 
    \,S_{T \rho}^{}\,g_T^{\perp}(x,\bm{k}_T^2) 
\\ &\quad \quad
    - \frac{\epsilon_T^{\alpha\rho} k_{T \rho}^{}}{M} \, 
    g^\perp(x,\bm{k}_T^2) \bigg] ,
    \label{eq:TMD_g2}
\end{split} 
\\
\begin{split} 
\Phi^{[i \sigma^{\alpha +}\gamma_5]}(x,\bm{k}_T)
& = S_T^\alpha \, h_1(x,\bm{k}_T^2)
  + S_L\,\frac{p_T^\alpha}{M} \, h_{1L}^\perp(x,\bm{k}_T^2)
\\ & \quad
 - \frac{p_T^\alpha p_T^\rho
     -\frac{1}{2}\,p_T^2\,g^{\alpha\rho}_T}{M^2}\, S_{T \rho} \, 
   h_{1T}^\perp(x,\bm{k}_T^2)
\\ & \quad 
- \frac{\epsilon_T^{\alpha\rho} p_{T \rho}}{M} \, 
   h_1^\perp (x,\bm{k}_T^2)
 \,,
\end{split} 
\end{align}
where $S_L = S^+ M/P^+$, and the transverse tensors $g^{\alpha\rho}_T$ and $\epsilon_T^{\alpha \rho}$ are defined as
\begin{align} 
 g^{\alpha\rho}_T
&= g^{\alpha\rho} - n_{+}^{\alpha} n_{-}^{\rho}
- n_{-}^{\alpha} n_{+}^{\rho}\, ,
\\
\epsilon_T^{\alpha \rho} &=\epsilon^{\alpha \rho \beta \sigma} (n_{+})_{\beta} (n_{-})_{\sigma} .
\end{align}  

For the $\bm{k}_T$-integrated distributions, we correspondingly have
\begin{align}
\Phi^{[\gamma^+ \gamma_5]}(x)
& = S_L \, g_{1L}(x)\, ,				\\
\Phi^{[i \sigma^{\alpha +}\gamma_5]}(x)
& = S_T^\alpha \, h_1(x)\, ,				\\ 
\Phi^{\alpha [\gamma^+ \gamma_5]}_\partial(x)
& = S_T^\alpha M  g_{1T}^{(1)}(x)\, ,			\\
\Phi^{[\gamma^\alpha \gamma_5]}(x)
&= \frac{M}{P^+} S_T^\alpha \, g_T(x)\, ,
\end{align} 
where for any TMD $f=f(x,\bm{k}_T^2)$ we define 
\begin{align}
  f^{(1)}(x,\bm{k}_T^2) &= \frac{\bm{k}_T^2}{2 M^2} f(x,\bm{k}_T^2) \ ,
\\
  f^{(1)}(x) &= \int d^2\bm{k}_T\, f^{(1)}(x,\bm{k}_T^2) \ .
\end{align}
To avoid confusion with the structure function $g_1$, here we use the
notation $g_{1L}$ also for the helicity-dependent PDF, 
contrary to what is used in some of the TMD
literature~\cite{Bacchetta:2006tn}.

The connection between the TMDs and the $A_i$ and $B_i$ amplitudes has
been worked out in detail in the Appendix of Ref.~\cite{Metz:2008ib}
for $v=n_-$.  In Appendix~\ref{a:v} we extend these results to a 
non-lightlike vector $v$.  We shall not repeat here the calculations
but only quote the results relevant for our discussion, namely
\begin{align}
\begin{split} 
 g_{1L}(x, {\bm k}_T^2)
 & =  \int d\sigma d\tau\,\delta(\tau-x\sigma+x^2 M^2+{\bm k}_T^2) \\
 &   \quad \times \Bigl(- A_6 - B_{11} - x B_{12})\\
 & \quad \quad - \frac{\sigma -2xM^2}{2M^2}(A_7 + x A_8)\Bigr)\, ,
\label{e:g1_AB}
\end{split}
\\
\begin{split}  
 g_{1T}(x, {\bm k}_T^2)
 & =  \int d\sigma d\tau\,\delta(\tau-x\sigma+x^2 M^2+{\bm k}_T^2) \\
 &  \quad \times \Bigl(A_7 + x A_8\Bigr)\, ,
\label{e:g1T_AB}
\end{split} 
\\
\begin{split} 
 g_T(x, {\bm k}_T^2)
 & =  \int d\sigma d\tau\,\delta(\tau-x\sigma+x^2 M^2+{\bm k}_T^2) 
\\ &\quad \times
   \Bigl( - A_6 - \frac{\tau-x \sigma +x^2 M^2}{2 M^2} A_8 \Bigr)\, .
   \label{e:gT_AB}
\end{split} 
\end{align} 
As anticipated, we see that $B_i$ terms appear also in the function
$g_{1L}$, which survives if the correlator is integrated over $\bm{k}_T$.

\subsection{Lorentz invariance relations}

From the preceding discussion, using the techniques discussed for example in
Ref.~\cite{Tangerman:1994bb}, it is possible to derive
the so-called Lorentz invariance relation (LIR) 
\begin{align} 
g_T(x) &= g_{1L}(x)
	+ \frac{d}{dx}\,g_{1T}^{(1)}(x)
	+ \widehat{g}_T(x)\, ,
\label{e:mod_LIR}
\end{align} 
where the function $\widehat{g}_T$ is given by
\begin{equation} 
\begin{split} 
\widehat{g}_T(x)
&= \int d^2 {\bm k}_T  d\sigma d\tau\,
   \delta(\tau-x\sigma+x^2 M^2+{\bm k}_T^2)\\
&  \quad \times
\Big[ B_{11} + x B_{12} 
- \frac{{\bm k}_T^2}{2 M^2}
  \Big( \frac{\partial A_7}{\partial x}
	+ x \frac{\partial A_8}{\partial x}
  \Big)
\Big]\\ 
 & \quad +\pi \int d\sigma d\tau\,
   \delta(\tau-x\sigma+x^2 M^2+{\bm k}_T^2)\,{\bm k}_T^2 \\
& \quad \quad \quad \times
\left. \frac{\sigma -2xM^2}{2M^2}
  \Bigl( A_7 + x A_8 \Bigr)
\right|_{{\bm k}_T^2 \to 0}^{{\bm k}_T^2 \to \infty}\ .
\label{e:gThat}
\end{split}
\end{equation} 
The proper operator definition for $\widehat{g}_T$ can be traced back
to Ref.~\cite{Bukhvostov:1983as} (see
also~\cite{Belitsky:1997ay,Kundu:2001pk}), and requires the introduction
of the twist-3 quark-gluon-quark correlator 
\begin{equation}
\begin{split}
&i \Phi_F^\alpha(x,x') =
 \int \frac{d \xi^- d \eta^-}{(2 \pi)^2}\,
e^{i k \cdot \xi}\,
e^{i (k'-k) \cdot \eta}\, \delta_T^{\alpha \rho}
\\
& \quad \times 
\langle P| \overline{\psi}(0)\,
 {\cal W}^v_{(0,\eta)} \, 
i g\, F^{+ \alpha}(\eta) \, 
 {\cal W}^v_{(\eta,\xi)} \, 
\psi(\xi) |P \rangle 
\Big|_{\substack{
\xi^+=\xi_T=0 \\
\eta^+=\eta_T=0}}
\\ &
\stackrel{\text{LC}}{=}
\int \frac{d \xi^- d \eta^-}{(2 \pi)^2}\,
e^{i k \cdot \xi}\,
e^{i (k'-k) \cdot \eta}
\\
& \quad \times 
\langle P| \overline{\psi}(0)\,
i g \, \partial^+_\eta A_T^\alpha(\eta) \, 
\psi(\xi) |P \rangle 
\Big|_{\substack{
\xi^+=\xi_T=0 \\
\eta^+=\eta_T=0}}\ ,
\label{e:phiF}
\end{split}  
\end{equation}
where $F^{+ \alpha}$ is the gluon field strength tensor, $k'$ is
the gluon momentum, and $x' = k'\cdot n_-/P\cdot n_-$.  Note that
this correlator has been discussed in slightly different forms in
Refs.~\cite{Boer:1997bw,Kanazawa:2000hz,Eguchi:2006mc,Boer:2003cm},
for example.
It can be expanded in terms of four scalar functions $G_F$,
$\widetilde{G}_F$, $H_F$ and $E_F$ according to
\cite{Boer:1997bw,Kanazawa:2000hz} 
\begin{equation}
\begin{split} 
i\Phi_F^\alpha(x,x')
&= \frac{M}{4}\biggl[ G_F(x,x') i \epsilon_T^{\alpha \rho} S_{T \rho}
 + \widetilde{G}_F(x,x') S_T^\alpha\gamma_5
\\
 & \quad
+ H_F(x,x')\, S_L \gamma_5 \gamma_T^\alpha
+ E_F(x,x')\, \gamma_T^\alpha \biggr] \nslash_+\ .
\label{e:phiFdecomp}
\end{split} 
\end{equation} 
Hermiticity and parity invariance impose that these functions are
real and either odd or even under the interchange of $x$ and $x'$
\cite{Kanazawa:2000hz},
\begin{align}
G_F(x,x') &= G_F(x',x)\, , 
&
\widetilde{G}_F(x,x') &= -\widetilde{G}_F(x',x)\, , \\
E_F(x,x') &= E_F(x',x)\, ,
&
H_F(x,x') &= -H_F(x',x)\, . 
\end{align} 
We can then express the function $\widehat{g}_T$ as
\begin{equation}
\begin{split} 
M S_T^\alpha\, \widehat{g}_T(x) &=
-\int d x'\,
 \frac{i\Phi_F^{\alpha [\gamma^+\gamma_5]}(x',x)}{(x-x')^2}
\\ & =
M S_T^\alpha\ {\cal P} \int d x'\,
 \frac{\widetilde{G}_F(x,x')/(x-x')}{x-x'},
\label{e:hatg}
\end{split} 
\end{equation} 
where ${\cal P}$ denotes the principal value integral.
(The need for the principal value was apparently overlooked in
Refs.~\cite{Belitsky:1997ay,Kundu:2001pk}.)
The imaginary part arising from the pole at $x=x'$ cannot give a
contribution to the LIR in Eq.~\eqref{e:mod_LIR}, but rather
contributes to a LIR involving the functions $f_T$ and
$f_{1T}^{\perp(1)}$, which we do not discuss here.
We note that $\widehat{g}_T$ is a ``pure twist-3'' function,
being part of the twist-3 correlator of Eq.~\eqref{e:phiF}.
Since the integrand in Eq.~\eqref{e:hatg} is antisymmetric in
$x \leftrightarrow x'$, one obtains the nontrivial property
\begin{equation}
   \int_0^1 dx\, \widehat{g}_T(x) = 0 \ .
\label{eq:hatsumrule}
\end{equation}

In some analyses \cite{Tangerman:1994bb,Boer:1997nt} $\widehat{g}_T$
was believed to vanish because
(i) the $B_i$ parton correlation functions were not taken into account,
(ii) the partial derivatives in Eq.~\eqref{e:gThat} were neglected since
an explicit $x$-dependence of the PCFs is generated only through the
additional $v$-dependence,
(iii) the boundary terms like the last terms in~\eqref{e:gThat} were
neglected.
However, {\it none} of these assumptions is justified, as we show
explicitly in a quark-target perturbative calculation in
Appendix~\ref{a:model}. 
We can further draw some model-independent conclusions about the
boundary terms by comparing them with the expression for $g_{1T}$
in Eq.~\eqref{e:g1T_AB}.
Positivity bounds imply that
$|\bm{k}_T^2 g_{1T}|\le M |\bm{k}_T| f_1$ \cite{Bacchetta:1999kz},
which is sufficient to guarantee that the $\bm{k}_T^2 = 0$ boundary
term indeed vanishes.
However, since $g_{1T}$ behaves as $1/\bm{k}_T^4$ at large $\bm{k}_T$
\cite{Bacchetta:2008xw}, the boundary term at $\bm{k}_T^2 =\infty$
cannot be neglected.

If $\widehat{g}_T$ is nonetheless neglected, it is possible to express
the twist-3 function $g_T$ in terms of the twist-2 functions $g_{1L}$
and $g_{1T}$~\cite{Mulders:1995dh,Tangerman:1994bb}.  
Relations of this kind have been often mistakenly called 
Lorentz invariance
relations~\cite{Mulders:1995dh,Tangerman:1994bb,Henneman:2001ev}, but 
should {\it not} be confused with the correct Lorentz invariance
relations such as in Eq.~(\ref{e:mod_LIR}).

In the literature, model calculations have been used to argue that
the pure twist-3 terms are not necessarily small
\cite{Jaffe:1990qh,Harindranath:1997qn}.  For example, $\widehat{g}_T$
can be computed perturbatively in the quark-target model of
Refs.~\cite{Harindranath:1997qn,Kundu:2001pk}.
Using Eqs.~(38), (40) and (42) of Ref.~\cite{Kundu:2001pk} one finds
\begin{align} 
g_T(x)-g_{1L}(x)
  & = \frac{\alpha_s}{2 \pi} C_F \ln{\frac{Q^2}{\mu^2}}
    \bigl[2x-\delta(1-x) \bigr]\, ,
\\
  g_{1T}^{(1)}(x) & = - \frac{\alpha_s}{2 \pi} C_F 
    \ln{\frac{Q^2}{\mu^2}}\,x (1-x)\, ,
\label{eq:harindranath}
\end{align} 
where $C_F=4/3$, $\mu$ is an infrared cutoff, 
and from Eq.~\eqref{e:mod_LIR} one has
\begin{align} 
\widehat{g}_T(x) &=
\frac{\alpha_s}{2 \pi}C_F \ln{\frac{Q^2}{\mu^2}}\,
\bigl[1-\delta(1-x)\bigr]\, .
\label{e:gThat_LIR}
\end{align}  
From this calculation one can see that $\widehat{g}_T$ is comparable to
the size of the other twist-2 functions.  Moreover, its lowest moment
vanishes, so that the nontrivial requirement of Eq.~\eqref{eq:hatsumrule}
is fulfilled.  In Appendix~\ref{a:model2} we confirm the above result
(for $x<1$ only) starting directly from the definition in
Eq.~\eqref{e:hatg}.

\subsection{Equations of motion relations}

The equations of motion (EOM) for quarks, $\Dslash\psi = m\psi$,
imply further relations between twist-2 and pure twist-3 functions
(namely, between $qq$ and $qgq$ matrix elements).  They are referred
to as ``equations of motion relations'', and for the case of
interest here are expressed as
\begin{align}
  g_{1T}^{(1)}(x)&= x g_T(x) - x \widetilde{g}_T(x) 
    - \frac{m}{M}\,h_1(x) \ ,
    \label{eq:EOM_gT_1} 
\end{align}
where
\begin{equation}
\begin{split} 
x M &S_T^{\sigma}\, \widetilde{g}_T(x)
= {\cal P} \int d x'\,
  \frac{i\Phi_{F \rho}^{[\gamma^+ \gamma_T^{\sigma} \gamma_T^{\rho}
			\gamma_5]}(x',x)}{x-x'}
\\ & =
M S_T^{\sigma}
\left( {\cal P} \int dx'\, \frac{G_F(x,x')}{2(x'-x)}
	      + \int dx'\, \frac{\widetilde{G}_F(x,x')}{2(x'-x)}
\right)\ .
\label{e:tildeg}
\end{split} 
\end{equation} 
The full list of EOM relations can be found in
Ref.~\cite{Bacchetta:2006tn}.

Using Eq.~\eqref{eq:EOM_gT_1} to eliminate $g_{1T}^{(1)}(x)$ in
Eq.~\eqref{e:mod_LIR}, one finds the differential equation
\begin{align}
x \frac{d}{dx}
\left( g_T - \widetilde g_T -\frac{m}{M}\frac{h_1}{x} \right)
+ g_{1L} - \widetilde g_T -\frac{m}{M}\frac{h_1}{x} + \widehat{g}_T
= 0\ .
\end{align}
Assuming that the relevant functions are integrable by
$\int_x^1 (dy/y)$ and solving for $g_T$ one finds
\begin{align}
\begin{split}
  g_T(x)
& = \int_x^1\frac{dy}{y} \Bigl(g_{1L}(y)+\widehat{g}_T(y)\Bigr)  \\
& \quad + \widetilde g_T^\star(x) + \frac{m}{M}(h_1/x)^\star(x)\ ,
\label{eq:WW_gT}
\end{split}
\end{align}
where we have introduced the shorthand notation
\begin{align}
f^\star(x) \equiv f(x) - \int_x^1\frac{dy}{y} f(y) 
           = -  \int_x^1\frac{dy}{y} \frac{d}{dy}\left[ yf(y) \right] \ .
\label{e:star}
\end{align}
Note that if the integrals over $x$ and $y$ can be exchanged,
the function $f$ satisfies
\begin{align}
  \int_0^1 dx\, f^\star(x) = 0\ .
\label{eq:intfstar}
\end{align}
In general, however, this is not necessarily true, as stressed in
Refs.~\cite{Jaffe:1989xx,Jaffe:1990qh}.

In DIS on a quark-target, $\widetilde{g}_T$ can be computed using
Eqs.~(38) and (43) of Ref.~\cite{Kundu:2001pk}, giving
\begin{equation} 
\begin{split} 
x g_T(x)-\frac{m}{M} h_1(x)
  & = \frac{\alpha_s}{2 \pi}C_F \ln{\frac{Q^2}{\mu^2}}
\\ &\quad  \times
    \biggl[-x(1-x)+\frac{\delta(1-x)}{2} \biggr]\, ,
\label{eq:harindranath2}
\end{split} 
\end{equation} 
and using Eq.~\eqref{eq:EOM_gT_1} we obtain
\begin{align} 
\widetilde{g}_T(x) &=
\frac{\alpha_s}{2 \pi} C_F \ln{\frac{Q^2}{\mu^2}}\,
\frac{\delta(1-x)}{2}\, .
\end{align}  
Again we see that the twist-3 function $\widetilde{g}_T$ has a size
comparable to that of the other twist-2 functions.

\subsection{Breaking of the Wandzura--Wilczek relation}

The hadronic tensor relevant for spin-dependent DIS structure functions
is given by the standard Lorentz decomposition
\begin{align}
\begin{split}
& W^{\mu\nu}(P,q) 
  = \frac{1}{P\cdot q} \varepsilon^{\mu\nu\rho\sigma} q_\rho \\
& \quad \times \Big[ S_\sigma g_1(x_B,Q^2) 
    + \Big( S_\sigma - \frac{S\cdot q}{P\cdot q}\, p_\sigma
	\Big) g_2(x_B,Q^2)
  \Big] \ ,
\label{eq:Wmunu}
\end{split}
\end{align}
where $q$ is the momentum of the exchanged photon and
$x_B=Q^2/(2P \cdot q)$ is the Bjorken variable.
In general the structure functions $g_1$ and $g_2$ in
Eq.~(\ref{eq:Wmunu}) are functions of the physical (external)
variable $x_B$ and are given by convolutions of the hard
$\gamma^*$--parton scattering coefficient functions and
the relevant PDFs.
At {\it leading order} in $\alpha_s$, and including terms up to twist~3,
they can be expressed in terms of the distributions $g_{1L}^a$ and
$g_T^a$ (where we now explicitly include the flavor index $a$)
introduced above as \cite{Bacchetta:2006tn}
\begin{align} 
g_1(x)&= \frac{1}{2}\,\sum_a e_a^2\;g_{1L}^a(x)\ \, ,
\label{e:g1}
\\
\label{e:g12}
g_1(x) + g_2(x)&= \frac{1}{2}\,\sum_a e_a^2\; g_T^a(x)\, ,
\end{align} 
where for simplicity we have suppressed the $Q^2$ dependence.
This then enables the difference between the full $g_2$ structure
function and the WW approximation \eqref{eq:g2WWdef} to be written as
\begin{equation} 
\begin{split} 
 g_2&(x) - g_2^{\rm WW}(x) 
\\
&= \frac{1}{2}\,\sum_a e_a^2
\biggl(
    \widetilde g_T^{a \star}(x) 
    + \frac{m}{M} (h_1^a/x)^\star(x) 
    + \int_x^1\frac{dy}{y} \widehat{g}_T^a(y) 
\Biggr)\ ,
\label{eq:WWrelation_mod}
\end{split} 
\end{equation} 
which represents the breaking of the WW relation.
Note that the right-hand-side of Eq.~\eqref{eq:WWrelation_mod}
contains a quark mass term and {\it two} pure twist-3 terms.
This is the main result of our analysis.

From Eq.~\eqref{eq:intfstar} the $x$ integral of the pure twist-3
function containing $\widetilde{g}^a_T$ and the mass term vanish.
Using Eq.~\eqref{eq:hatsumrule}, and assuming that $\widehat{g}^a_T$
is regular enough to exchange the $x$ and $y$ integrals, we see that
the $\widehat{g}^a_T$ term also vanishes.
This implies that the above expression for $g_2$ satisfies the
Burkhardt--Cottingham sum rule, Eq.~(\ref{eq:BC}), which is not in
general guaranteed in the OPE~\cite{Jaffe:1989xx,Jaffe:1990qh}.  

To obtain the WW relation one must neglect quark mass terms
compared to the hadron mass (which can be reasonably done for light
quarks), and either neglect both of the pure twist-3 terms, or assume
that they cancel each other. 
The explicit quark-target perturbative calculations show that such a
cancellation does not take place in general, and that the size of the
WW breaking term can be comparable to the size of $g_2^{\rm WW}$,
\begin{align}
\begin{split} 
  &g_2^{\rm WW}(x)  = 1 -\delta(1-x)
-\frac{\alpha_s}{2 \pi}C_F \ln{\frac{Q^2}{\mu^2}}
\\ & \hspace{-0.5cm}\times
\biggl[-\log \frac{(1-x)^2}{x} + \frac{3}{2}\,\delta(1-x)
	+ \frac{2 x^2}{(1-x)_+} + \frac{1}{2}\biggr],
\label{eq:g2WWq}
\end{split}  
\\
\begin{split} 
  g_2(x)-&g_2^{\rm WW}(x) = \delta(1-x) -1 
+\frac{\alpha_s}{2 \pi}C_F \ln{\frac{Q^2}{\mu^2}}
\\ & \hspace{-0.5cm}\times
\biggl[-\log \frac{(1-x)^2}{x} +\frac{1}{2}\,\delta(1-x)
	+ \frac{2}{(1-x)_+} - \frac{3}{2}
\biggr] .
\end{split} 
\end{align}
To obtain the above expressions we again made use of the results
in Ref.~\cite{Kundu:2001pk}. 
Note that both $g_2^{\rm WW}$ and the total $g_2$ structure function
in the quark-target model respect the BC sum rule.

\section{Constraints from data}    
\label{sec:expWW}

It is often stated in the literature (see {\it e.g.}
Ref.~\cite{Avakian:2007mv}) that the WW relation holds experimentally
to a good accuracy.  While there are certainly indications
that this may indeed be so \cite{Anthony:2002hy,Amarian:2003jy}, it
is important to quantify the degree to which this relation holds and
place limits on the size of its violation.  This is the focus of this
section.

We define the experimental WW breaking term $\Delta_{\rm ex}(x_B)$
as the difference between the experimental data and $g_2^{\rm WW}$,
\begin{align}
  \Delta_{\rm ex}(x_B,Q^2)
  = g_2^{\rm{ex}}(x_B,Q^2) - g_2^{\rm{WW}}(x_B,Q^2) \ ,
\end{align} 
with the Wandzura--Wilczek term computed using the LSS2006 (set 1)
fit of the $g_1$ structure function \cite{Leader:2006xc}.  The fit
was performed including a phenomenological higher-twist term and
target mass corrections in order to extract the pure twist-2
contribution, $g_1^{\rm LT}$.  Using parametrizations of $g_1$
which do not account for the $1/Q$ power corrections
\cite{deFlorian:2008mr,Hirai:2006sr} would risk inadvertantly including
spurious higher twist contributions when computing the WW approximation.
We will demonstrate the impact of this difference by comparing our
$g_2^{\rm WW}$ with $(g_2^{\rm WW})'$ computed using the total $g_1$
instead of $g_1^{\rm LT}$ in Eq.~\eqref{eq:g2WWdef}.

For proton targets we consider data from the SLAC E142 \cite{Abe:1998wq}
and E155x \cite{Anthony:2002hy} experiments, while for the neutron only
the high-precision data sets from the SLAC E155x \cite{Anthony:2002hy},
and Jefferson Lab E99-117 \cite{Zheng:2004ce} and E01-012
\cite{Kramer:2005qe} experiments, obtained using $^2$H or $^3$He targets,
are included.
We checked explicitly that including the lower-precision data sets from
Refs.~\cite{Abe:1997qk,Abe:1998wq,Anthony:1996mw} does not alter the
fit results, except for artificially lowering the $\chi^2$ values due
to the much larger errors compared to the higher-precision data sets.
In total, there are 52 data points for the proton and 18 points for
the neutron, which are used separately to fit the WW breaking term
$\Delta$.  Systematic errors, when quoted, were added in quadrature.
For the shape of $\Delta$ we choose the form 
\begin{equation} 
\Delta(x_B,\alpha,\beta)
= \alpha (1-x_B)^\beta \bigl((\beta+2)x_B - 1\bigr)\, ,
\end{equation} 
which vanishes at $x_B=1$, has no divergences at $x_B=0$, fulfills the
BC sum rule, and only has a single node.  We do not consider its $Q^2$
QCD evolution.  The evolution of $g_2$ has been studied numerically in
Ref.~\cite{Stratmann:1993aw} in the limit of a large number of colors.
Most of the data considered lie in the range
1~GeV$^2 \le Q^2 \le$ 10~GeV$^2$ where the effect of QCD evolution
is rather mild, as indicated also by the results of the E01-012
experiment \cite{Kramer:2005qe}.

The goodness of the fit is estimated using the $\chi^2$ function 
\begin{equation}
\chi^2=\sum_{i=1}^N
  \frac{\bigl[\Delta(x_{B i}) - \Delta_{\rm ex}(x_{B i})\bigr]^2}
{\sigma_{\rm ex}^2(x_{B i})}\, .
\end{equation} 
To quantify the size of the breaking term $\Delta$ compared to
$g_2^{\rm WW}$ we define, for any interval
$[x_B^{\rm{min}}, x_B^{\rm{max}}]$, the ratio of their quadratic
integrals
\begin{align}
r^2 = \frac{\int_{y^{\rm{min}}}^{y^{\rm{max}}} dy\, x_B^2\Delta^2(x_B)}
           {\int_{y^{\rm{min}}}^{y^{\rm{max}}} dy\, x_B^2g_2^2(x_B)} \ ,
\end{align}
with $y=\log(x_B)$.   The value of $r$ is a good indicator of the
relative magnitude of $\Delta$ and $g_2$, which change sign as a function
of $x_B$.  In practice we compute $r$ at the average kinematics of the
E155 experiment \cite{Anthony:2002hy}.
For the proton, we consider three intervals: the entire measured $x_B$
range, [0.02,1]; the low-$x_B$ region, [0.02,0.15]; and the high-$x_B$
region, [0.15,1].  For the neutron, due to the limited statistical
significance of the low-$x_B$ data, we limit ourselves to quoting the
value of $r$ for the large-$x_B$ region, [0.15,1].

\begin{table*}[ht]
\centering
\begin{tabular}[c]{crlcccc}
\hline & \multicolumn{2}{l}{\bf proton} & $\chi^2$/d.o.f. 
& $r_\text{tot}$ & $r_\text{low}$ & $r_\text{hi}$ \\\hline
  (I)  & $\Delta$ & = 0 
    & 1.22 \hspace*{0cm} \\
  (II)  
    & $\Delta$ & = $\alpha (1-x_B)^\beta \bigl((\beta+2) x_B-1\bigr)$
                                        \hspace*{.1cm}\\
    & $\alpha$ & = $0.13 \pm 0.05$ \\
    & $\beta$ & = $4.4 \pm 1.0$ 
    & 1.05  &  15--32\% & 18--36\% & 14--31\% \\
\hline & \multicolumn{2}{l}{\bf neutron} \\\hline
  (I)   & $\Delta$ & = 0 
    & 1.66 \hspace*{0cm} \\
  (II)  
    & $\Delta$ & = $\alpha (1-x_B)^\beta \bigl((\beta+2) x_B-1\bigr)$
                                        \hspace*{.1cm}\\
    & $\alpha$ & = $0.64 \pm 0.92$ \\
    & $\beta$ & = $24 \pm 10$ 
    & 1.11  &   &  & 18--40\% \\
\hline
\end{tabular}
\caption{Results of the 1-parameter fits of the WW breaking term
  $\Delta$ for different choices of its functional form.  The value
  $r$ of the relative size of the breaking term is computed for three
  regions of $x_B$: the entire measured $x_B$ range, [0.02,1]; the
  low-$x_B$ region, [0.02,0.15]; and the high-$x_B$ region, [0.15,1]. 
  See text for further details.} 
\label{tab:WWfits}
\end{table*}

\begin{figure*}
  \centering
  \includegraphics[width=0.49\linewidth]{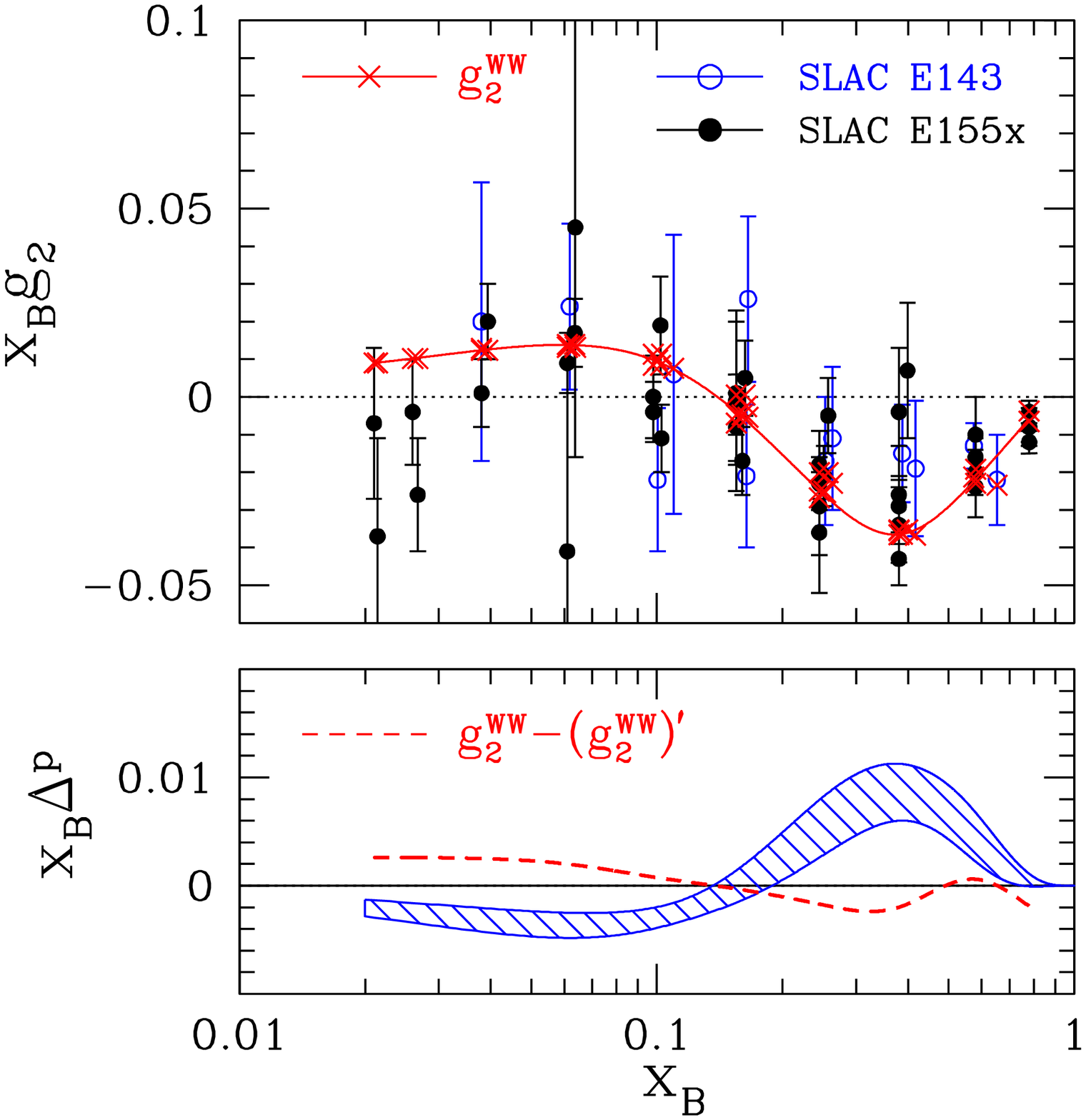}
  \includegraphics[width=0.49\linewidth]{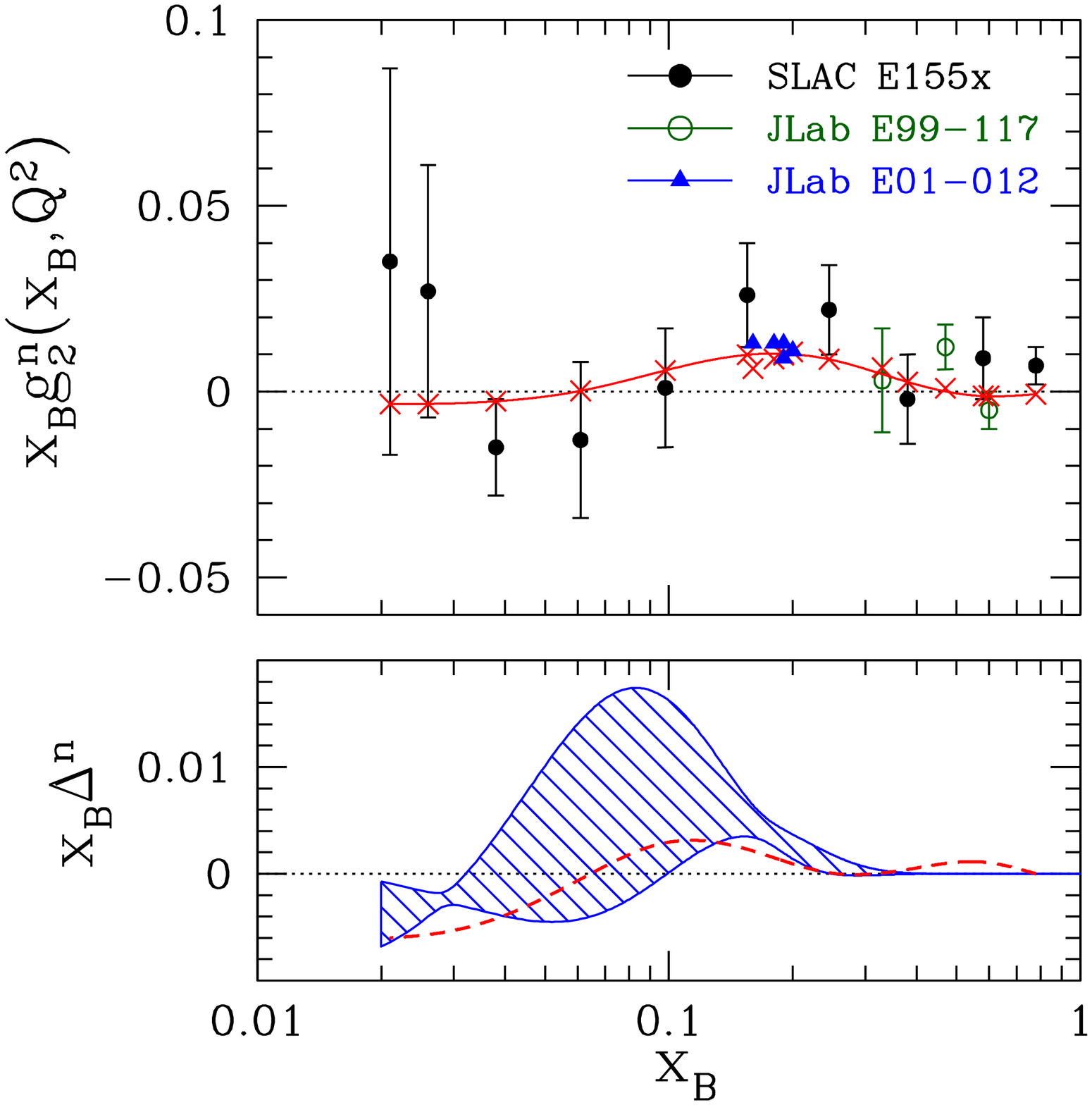}
  \caption{
    {\it Top panels}: 
    Experimental proton and neutron $g_2$ structure functions compared to
    $g_2^{\rm WW}$. The crosses represent $g_2^{\rm WW}$ computed at the
    experimental kinematics, while the solid lines are $g_2^{\rm WW}$
    computed at the average $Q^2$ of the E155x experiment.
    Data points for the proton target \cite{Abe:1998wq,Anthony:2002hy}
    have been slightly shifted in $x_B$ for clarity. For the neutron
    only the high-precision data from
    \cite{Anthony:2002hy,Zheng:2004ce,Kramer:2005qe} are included. 
    {\it Bottom panels}:
    The WW-breaking term $\Delta$ fitted to $\Delta_{\rm ex}$ 
    computed using the LSS2006 $g_!^{\rm LT}$
    (hashed region). The dashed line represents $g_2^{\rm WW}-(g_2^{\rm WW})'$,
    the spurious HT contribution to $\Delta$ that would be obtained using
    the total $g_1$ to compute $\Delta_{\rm ex}$.
  }
  \label{fig:g2WW}
\end{figure*}

The results of the fits are presented in Table~\ref{tab:WWfits} and
Figure~\ref{fig:g2WW}.
The proton fit displays a positive WW breaking at large-$x_B$ and a
negative breaking at small-$x_B$. The size of the breaking term is 
typically 15--35\% of the size of $g_2$ (see the $r$ values in
Table~\ref{tab:WWfits}).
The neutron fit is completely dominated by the high-precision
E01-012 data, which are concentrated on a very limited $x_B$ range;
it clearly indicates a 18--40\% breaking of the WW relation at high $x_B$,
but cannot be used to conclude much at lower $x_B$ values.
A striking feature of the proton WW-breaking term in Fig.~\ref{fig:g2WW}
is that it is comparable in size and {\it opposite} in sign to
$g_2^{\rm WW}-(g_2^{\rm WW})'$.  It is essential, therefore, to use
fits of $g_1$ that subtract higher twist terms, which would otherwise
largely cancel the proton WW-breaking term and obscure the violation
of the WW relation.
In the case of the neutron one would generally obtain an enhancement
of the WW-breaking term, although the experimental uncertainties there
are considerably larger.

In summary, we have found that the experimental data are consistent
with a substantial breaking of the WW relation (\ref{eq:WWrel}).
Previous analyses have verified the WW relation only qualitatively,
and using parametrizations which do not subtract higher twist terms in
$g_1$.  The present analysis clearly demonstrates that this can give
the misleading impression that the WW relation holds to much better
accuracy than it does in more complete analyses where the higher twist  
corrections have been consistently taken into account.
More data are certainly needed to pin down the breaking of the WW
relation to higher precision.  New data are expected soon from the
HERMES Collaboration and from the d2n (E06-014) and SANE (E07-003)
experiments at Jefferson Lab \cite{e06-014,e07-003}.

\section{Toward a deeper understanding of quark-gluon-quark correlations}  
 
In the past, since the LIR-breaking $\widehat{g}_T$ term was not 
considered in Eq.~\eqref{eq:WWrelation_mod} and the quark-mass term with 
$h_1$ was neglected, the breaking of the WW relation was considered to 
be a direct measurement of the pure twist-3 term $\widetilde{g}_T$. 
The presumed experimental validity of the WW relation was therefore
taken as evidence that $\widetilde{g}_T$ is small.  This observation
was then generalized to assume that all pure twist-3 terms are small.
In contrast, the present analysis shows that, precisely due to the
presence of $\widehat{g}_T$, the measurement of the breaking of the WW
relation does {\it not} provide information on a single pure twist-3
matrix element.  Even if in future the WW relation were to be found to be
satisfied to greater accuracy than the present data suggest, one could
only conclude that the sum of the terms in (\ref{eq:WWrelation_mod}) is
small,
\begin{align}
\sum_a e_a^2
\biggl(
- \widetilde g_T^a(x)
+ \int_x^1 \frac{dy}{y}
  \Bigl(\widehat{g}_T^a(y)+\widetilde g_T^a(y) \Bigr)
\biggr)
%
\approx 0 \ .
\label{eq:WWexper}
\end{align}
This can occur either because $\widehat{g}_T^a$ and $\widetilde g_T^a$
are both small, or because they (accidentally) cancel each other.
No information can be obtained on the size of the twist-3
quark-gluon-quark term $\widetilde g_T$ from the experimental data
on $g_2$ alone.
Note that these results were essentially already obtained in
Ref.~\cite{Metz:2008ib}.  In that work, however, the authors considered
the WW breaking to be small and {\it assumed} that $\widetilde{g}_T^a$
was small (which we argue is not necessarily the case), concluding that
$\widehat{g}_T^a$ is also small.

Of course it is desirable to test our conclusions empirically.
A reliable way to investigate $\widetilde{g}_T$ experimentally is
through measurement of the function $g_{1T}^{(1)}$.  This function
is accessible in semi-inclusive deep inelastic scattering with
transversely polarized targets and longitudinally polarized lepton beams
(see, {\it e.g.}, the second line of Tab.~IV in Ref.~\cite{Boer:1997nt}).
Preliminary data related to this function have been presented by the
COMPASS Collaboration~\cite{Parsamyan:2007ju} and more are expected from
the HERMES Collaboration and from the E06-010 experiment at Jefferson Lab
\cite{e06-010}.  Using the EOM relation~\eqref{eq:EOM_gT_1} and assuming
$m=0$, one obtains
\begin{align}
  x \widetilde{g}_T(x) &= x g_T(x) - g_{1T}^{(1)}(x) \ .
\end{align}
In combination with the measurement of the WW breaking, this can be
used to determine the size of twist-3 function $\widehat g_T$.
(Alternatively, one can use the LIR \eqref{e:mod_LIR}.)

The importance of separately studying $\widetilde{g}_T$ and
$\widehat{g}_T$ resides in the fact that these are projections of
different combinations of the twist-3 functions $G_F(x,x')$ and
$\widetilde{G}_F(x,x')$.  As with all other terms in the decomposition
of the quark-gluon-quark correlator in Eq.~\eqref{e:phiFdecomp}, these
functions are involved in the evolution equation of twist-3 collinear
PDFs~\cite{Balitsky:1987bk,Belitsky:1997zw}, in the evolution of the
transverse moments of the TMDs~\cite{Kang:2008ey,Vogelsang:2009pj},  
in the calculation of processes at high transverse momentum
\cite{Eguchi:2006mc}, and in the calculation of the high transverse
momentum tails of TMDs~\cite{Ji:2006ub,Koike:2007dg}.  Ultimately,
through a global study of all of these observables, one could 
simultaneously obtain better knowledge of twist-3 collinear functions
and twist-2 TMDs, and at the same time test the validity of the
formalism.  Gathering as much information as one can on the
quark-gluon-quark correlator is essential to reach this goal.
The separation of the functions $\widetilde{g}_T$ and $\widehat{g}_T$
is an important first step in this direction.

\section{Conclusions}

In this analysis we have shown that the Wandzura--Wilczek relation
for the $g_2$ structure function is violated by a quark mass term, and
two distinct pure twist-3 contributions, containing
the parton distribution functions $\widehat{g}_T$ and $\widetilde{g}_T$.
As evident from their definitions in Eqs.~\eqref{e:hatg} and
\eqref{e:tildeg} respectively, these correspond to two different
projections of the general quark-gluon-quark correlator in
Eq.~\eqref{e:phiF}.  Their measurement can give unique and complementary 
information on twist-3 physics.

The two twist-3 functions have some interesting connections with the
formalism of transverse momentum distributions.  One of them is involved
in the equation-of-motion relation expressed in Eq.~\eqref{eq:EOM_gT_1},
while the other is involved in the Lorentz invariance relation in
Eq.~\eqref{e:mod_LIR}.  Both relations contain the same moment of the
transverse momentum distribution $g_{1T}$.  From the theoretical point
of view, this is another intriguing example of the interplay between
transverse momentum distributions and (collinear) twist-3 distributions.
From the phenomenological point of view, this means that a measurement
of the function $g_{1T}$ in semi-inclusive DIS in principle allows one
to separately measure $\widehat{g}_T$ and $\widetilde{g}_T$.

Although the Wandzura--Wilczek relation is often used to simplify the
treatment of twist-3 and TMD physics, we stress that there are no
compelling theoretical or phenomenological grounds supporting its
validity.  In fact, using the experimental information currently
available, we were able to provide a quantitative assessment of the
violation of the Wandzura--Wilczek relation.  Assuming a simple
functional form for the WW-breaking term, we found that it can be
as large as 15--40\% at the 1-$\sigma$ confidence level.

As new data become available, it should be possible to better pin down
the violation of the Wandzura--Wilczek relation and measure the
transverse momentum distribution $g_{1T}$ in semi-inclusive DIS.
This will offer us a deeper look into the physics of quark-gluon-quark
correlations and its connection to transverse momentum distributions.


\begin{acknowledgments}
We are grateful to M. Burkardt and A.~Metz for helpful discussions.
This work was supported by the DOE contract No. DE-AC05-06OR23177,
under which Jefferson Science Associates, LLC operates Jefferson Lab,
and NSF award No.~0653508.
\end{acknowledgments}

\appendix

\section{TMDs with a non-lightlike Wilson line direction} 
\label{a:v}
 
Factorization theorems beyond tree-level
\cite{Collins:1989gx,Ji:2004wu,Ji:2004xq,Collins:2004nx,Collins:2007ph}
demand a slightly non-lightlike vector $v$ in order to regularize the
lightcone (or rapidity) divergences~\cite{Collins:2003fm,Collins:2008ht}.  
In Ref.~\cite{Ji:2004wu}  the Wilson line vector is chosen to be timelike
and a parameter $\zeta^2=4(P\cdot v)^2/v^2$ is used as a regulator,
with the requirement that $\zeta^2 \gg M^2, \bm{k}_T^2$. In other
articles in the literature $v$ has been chosen to be
spacelike~\cite{Collins:1981uk}.

In addition to $k\cdot P$, $k^2$, $P\cdot v$ and $k\cdot v$, the PCFs
$A_i$ and $B_i$ can now in principle depend also on $v^2$.
We can derive the following relation between the invariants
\begin{equation}
 \frac{k\cdot v}{P \cdot v}= a x+\frac{2\sigma}{\zeta^2(1+a)}\, ,
\label{e:xazeta}
\end{equation}
with $a=\sqrt{1-4M^2/\zeta^2}$.  Neglecting terms of order $M^2/\zeta^2$
and $\sigma/\zeta^2$, the above expression reduces to $x$.
We therefore conclude that the PCFs depend on $\sigma,\tau,x$ and
additionally on $\zeta^2$.  To be precise, the definition of parton
correlation functions in \cite{Collins:2007ph} involves an additional
soft factor which is not included in the correlator $\Phi$.
The inclusion of the soft factor leads to an additional dependence on
a gluon rapidity parameter.  However, we leave this soft factor aside
since it plays no role in our subsequent discussion.

The expressions for the TMDs in Eqs.~\eqref{e:g1_AB},
\eqref{e:g1T_AB} and \eqref{e:gT_AB} then become
\begin{equation} 
\begin{split} 
 g_{1L}(x, &{\bm k}_T^2,\zeta^2)  =  \int d\sigma
 d\tau\,\delta(\tau-x\sigma+x^2 M^2+{\bm k}_T^2)\\ 
 &  \quad \times \Bigl[-A_{6}-a\Bigl(B_{11}+x B_{12}+\frac{4M^2}{\zeta^2(1+a)}B_{14}\Bigr)\\
 &  \quad \quad - \frac{\sigma -2xM^2}{2M^2}\Bigl(A_{7}+x
 A_{8}+\frac{4M^2}{\zeta^2(1+a)}B_{13}\Bigr)\Bigr],
\label{e:g1_AB_zeta}
\end{split} 
\end{equation} 
\begin{align} 
\begin{split} 
 g_{1T}(x, {\bm k}_T^2,\zeta^2) & =  \int d\sigma
 d\tau\,\delta(\tau-x\sigma+x^2 M^2+{\bm k}_T^2)\\ 
 &  \quad \times \Bigl[A_{7}+x
 A_{8}+\frac{4M^2}{\zeta^2(1+a)}B_{13}\Bigr],\label{e:g1T_AB_zeta} 
\end{split} 
\\
\begin{split} 
 g_{T}(x, {\bm k}_T^2,\zeta^2) & =  \int d\sigma
 d\tau\,\delta(\tau-x\sigma+x^2 M^2+{\bm k}_T^2) 
\\ &\quad \times \Bigl[-A_6-\frac{\tau-x \sigma +x^2 M^2}{2 M^2}A_8\Bigr],
   \label{e:gT_AB_zeta}
\end{split} 
\end{align}
The full expression for $\widehat{g}_T$ which generalizes
Eq.~\eqref{e:gThat} then becomes
\begin{equation} 
\begin{split} 
\widehat{g}&_T(x)  = \int d^2 {\bm k}_T\,  d\sigma d\tau\,
		     \delta(\tau-x\sigma+x^2 M^2+{\bm k}_T^2)\\
 &   \times \Big[B_{11} + xB_{12} + \frac{4M^2}{\zeta^2 (1+a)}B_{14}\\
 &   \quad -\frac{{\bm k}_T^2}{2 M^2}\Big(\frac{\partial A_7}{\partial x} + x \frac{\partial A_8}{\partial x} + \frac{4M^2}{\zeta^2 (1+a)}\frac{\partial B_{13}}{\partial x}\Big)\Big]\\ 
 &   +\pi \int d\sigma d\tau \,\delta(\tau-x\sigma+x^2 M^2+{\bm k}_T^2)\,{\bm k}_T^2 \\
 &   \quad \times \frac{\sigma -2xM^2}{2M^2} \Bigl(A_{7}+x A_{8}+\frac{4M^2}{\zeta^2(1+a)}B_{13})\Bigr)\Big|_{{\bm k}_T^2\rightarrow 0}^{{\bm k}_T^2\rightarrow \infty}.\label{e:gThat_zeta}
\end{split}
\end{equation}

\section{Parton correlation functions for a quark target} 
\label{a:model}

\begin{figure*}
  \includegraphics[height=2cm]{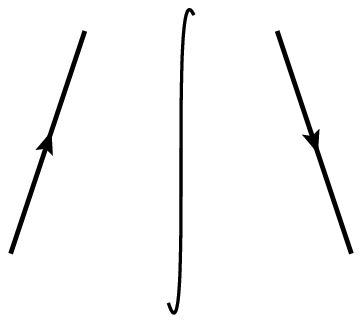}
  \qquad \quad
  \includegraphics[height=2cm]{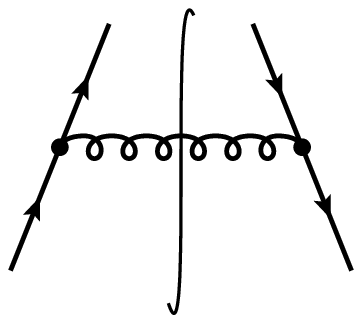}
  \qquad \quad
  \includegraphics[height=2cm]{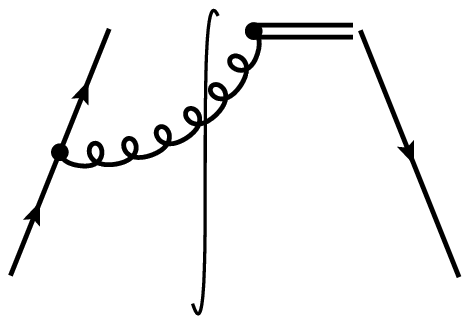}
  \qquad \quad
  \includegraphics[height=2cm]{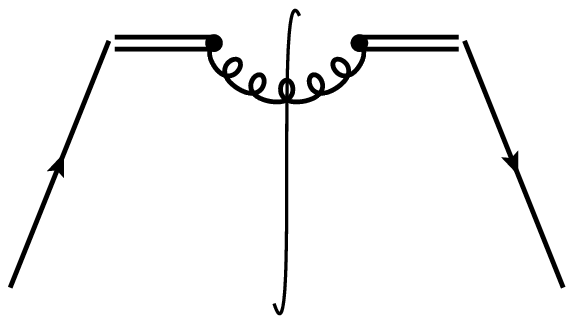}
 \caption{Diagrams in the quark-target calculation involving only
   real gluons.  The Hermitean conjugate diagrams, which are not
   shown, are also taken into account in the calculation.}
  \label{f:QTreal}
\end{figure*}

\begin{figure*}
  \includegraphics[height=2cm]{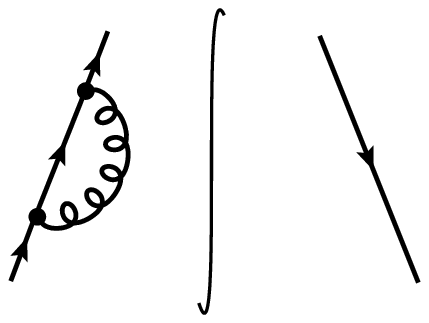}
  \qquad \quad
  \includegraphics[height=2cm]{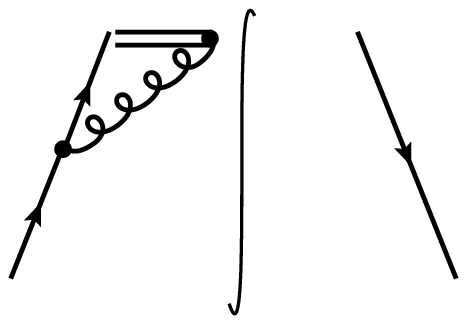}
  \qquad \quad
  \includegraphics[height=2cm]{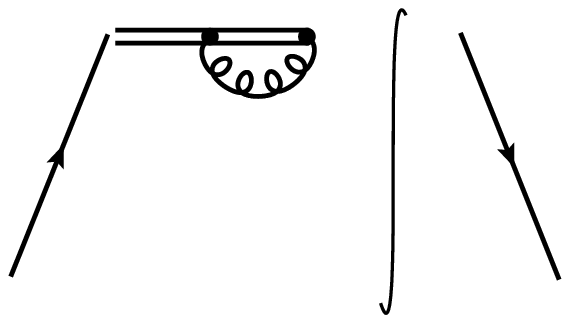}
 \caption{As in Fig.~\ref{f:QTreal} but for diagrams involving
   virtual gluons.}
  \label{f:QTvir}
\end{figure*}

In this Appendix we compute the parton correlation functions relevant
for our discussion of the WW relation for the case of a point-like
quark target.  The calculations are performed in the first non-trivial
order in perturbative QCD ({\i.e.}, at order $\alpha_s$) 
\cite{Harindranath:1997qn,Kundu:2001pk}.  To this end we insert
a complete set of intermediate states into Eq.~(\ref{e:corr1}).
To order $\alpha_s$, only the vacuum state and a one-gluon state
are relevant.  The involved Feynman diagrams are shown in 
Fig.~\ref{f:QTreal} (real gluon contributions) and Fig.~\ref{f:QTvir}
(virtual gluon contributions).

The correlator may be written as
\begin{equation}
\begin{split}
\Phi_{ij}(k,P,S; v) & =
\delta^{(4)}(P-k)\Phi_{ij}^{\rm{vir}}(m^2,\lambda^2,\zeta^2,\mu_R^2)
\\
& \quad +\Phi_{ij}^{\rm{real}}(k,P,S; v)\, ,
\label{e:RealVirtual}
\end{split}
\end{equation}
where $\Phi^{\rm{vir}}$ denotes the contributions from the vacuum
intermediate state.  Its kinematics is totally determined by the
four-dimensional delta-function $\delta^{(4)}(P-k)$ and depends
only on the quark mass $m$, with a small gluon mass $\lambda$ serving
here as an infrared regulator, and the parameter
$\zeta^2=4(P\cdot v)^2/v^2$ which regulates lightcone divergences.
By applying a renormalization procedure we can subtract ultra-violet
divergences in $\Phi^{\rm vir}$, which introduces a dependence on the
renormalization point $\mu_R^2$.
The virtual corrections can be written as
\begin{equation}
\begin{split}
\Phi_{ij}^{\rm{vir}}(k,P,S; v) & =  \delta^{(4)}(P-k)\langle
P,S,d|\, \bar{\psi}_j(0)\,{\cal W}^{v}_{(0,\infty)} \,|0\rangle
\\ &\quad \times \langle
0 |\,{\cal W}^{v}_{(\infty,0)} \,\psi_i(0)\,|P,S,d\rangle\, ,
\label{e:Vir}
\end{split}
\end{equation}
where the incoming on-shell quark is described by the state
$|P,S,d\rangle$, with $d$ a color index of the quark in the
fundamental SU(3) representation.
For the sake of brevity we will omit the explicit dependence
on and summation over the color indices in the following. 
Since we work in Feynman gauge, possible contributions from gauge
links at lightcone infinity are irrelevant~\cite{Belitsky:2002sm}.

The second contribution in Eq.~(\ref{e:RealVirtual}) is generated
by one gluon in the intermediate state.
To order $\alpha_s$ it is given by
\begin{equation}
\begin{split}
\Phi_{ij}^{\rm{real}}&(k,P,S; v)  =
\frac{1}{(2\pi)3}\sum_{\sigma,\beta} \delta^+((P-k)^2-\lambda^2)
\\ & \quad \times
\overline{M}^{\sigma,\beta}_j(k,P,S; v)\,
	  M^{\sigma,\beta}_i(k,P,S; v)\, ,
\label{e:Real}
\end{split}
\end{equation}
with $\overline{M}\equiv M^\dagger \gamma^0$,
$\delta^+(a^2)\equiv \delta(a^2)\Theta(a^0)$, $\sigma$ denotes the
polarization of the gluon in the intermediate state, and $\beta$ is
its color index in the adjoint representation of SU(3).  The matrix
element $M$ is then represented by
\begin{equation}
\begin{split}
M^{\sigma,\beta}_i&(k,P,S;v) =  \langle P-k,\sigma,\beta |
\psi_{i}(0) |P,S,d\rangle
\\ & +ig\int_0^\infty
d\lambda\, \langle P-k,\sigma,\beta | v\cdot A(\lambda v)\,\psi_{i}(0)
|P,S,d\rangle\, ,
\label{e:MReal}
\end{split}
\end{equation}
where $|P-k,\sigma,\beta \rangle$ denotes the intermediate gluon
state with a color index $\beta$. 
The leading perturbative contribution in $\alpha_s$ to the
matrix element $M$ gives
\begin{equation}
\begin{split}
M_i^{\sigma,\beta}&(k,P,S;v)  =
- gt^\beta\biggl(\frac{(\kslash+m)
  \epsslash_{\sigma}^*(P-k)}{[k^2-m^2+i\epsilon]}
\\ & \quad+ \frac{v\cdot \varepsilon^*_{\sigma}(P-k)}
		 {[v\cdot (P-k)+i\epsilon]}\biggr)_{il} u_l(P,S)\, ,
\label{e:Mpert}
\end{split}
\end{equation}
where $\varepsilon(P-k)$ denotes the gluon polarization vector and
$u$ is the quark spinor.  The color flow is given by the color matrix
$t^\beta$ in the fundamental representation.
Inserting~(\ref{e:Mpert}) into~(\ref{e:Real}) then yields
\begin{equation} 
\begin{split} 
\Phi^{\rm{real}}_{ij}&(k,P,S;v)  =  -\frac{\alpha_s}{(2\pi)^2}
C_F \delta^+((P-k)^2-\lambda^2)
\\ & \times
\Bigg[\frac{(\kslash+m)\gamma_\mu(\Pslash+m)(1+\gamma_5
\Sslash)\gamma^\mu(\kslash+m)}{[k^2-m^2+i\epsilon][k^2-m^2-i\epsilon]}
\\ & \quad
+\frac{(\Pslash+m)(1+\gamma_5\Sslash)\vslash(\kslash+m)}{[k^2-m^2-i\epsilon][v\cdot(P-k)+i\epsilon]}
\\ & \quad
+\frac{(\kslash+m)\vslash(\Pslash+m)(1+\gamma_5\Sslash)}{[k^2-m^2+i\epsilon][v\cdot(P-k)-i\epsilon]}
\\ & \quad
+\frac{v^2(\Pslash+m)(1+\gamma_5\Sslash)}{[v\cdot(P-k)+i\epsilon][v\cdot(P-k)-i\epsilon]}\Bigg]_{ij}\, .
\label{e:PhiReal}
\end{split} 
\end{equation} 
The various parton correlation functions in Eq.~\eqref{eq:decomp} can
be extracted from Eq.~(\ref{e:PhiReal}) by decomposing the numerators
in terms of the basis matrices $1$, $\gamma_5$, $\gamma^\mu$,
$\gamma^\mu\gamma_5$ and $\sigma^{\mu\nu}$.  In this way we obtain
expressions for parton correlation functions at leading order in
$\alpha_s$ for a quark target.
In the following we list only the PCFs $A_{6-8}$ and $B_{11-14}$ which
are relevant for the discussion of the Wandzura--Wilczek relation,
{\it cf.} Eqs.~(\ref{e:g1_AB})--(\ref{e:gT_AB}).  Setting
$a=\sqrt{\smash[b]{1-4 m^2/\zeta^2}}$, we find (to order $\alpha_s$)
\begin{equation} 
\begin{split} 
&A^{\rm{real}}_6(\tau,\sigma,x,\zeta^2) = \frac{C_F\alpha_s}{2
\pi^2}\delta^+(\tau-\sigma+m^2-\lambda^2) 
\\ 
 & \quad \times 
\Biggl[\frac{\tau+m^2}{\bigl(\tau-m^2\bigr)^2}+\frac{(1+a)(1+ax)+2\sigma/\zeta^2}{\bigl[\tau-m^2\bigr]\bigl[(1+a)(1-ax)-2\sigma/\zeta^2\bigr]}
\\
& \quad
+\frac{2(1+a)^2}{\bigl[(1-ax)^2(1+a)^2\zeta^2-4\sigma(1-ax)(1+a)+4\sigma^2/\zeta^2\bigr]}\Biggr],
\label{e:A6}
\end{split} 
\end{equation} 

\begin{align} 
A^{\rm{real}}_7&(\tau,\sigma,x,\zeta^2) =
0,
\label{e:A7}
\\
\begin{split} 
A^{\rm{real}}_8&(\tau,\sigma,x,\zeta^2) =  \frac{C_F\alpha_s}{2
\pi^2}\delta^+(\tau-\sigma+m^2-\lambda^2)
\\ & \times
\Biggl[\frac{-2m^2}{\bigl(\tau-m^2\bigr)^2}\Biggr],
\label{e:A8}
\end{split} 
\\
\begin{split} 
B^{\rm{real}}_{11}&(\tau,\sigma,x,\zeta^2) = \frac{C_F\alpha_s}{2
\pi^2}\delta^+(\tau-\sigma+m^2-\lambda^2)
\\ &\times
\Biggl[\frac{-(1+a)}{\bigl[\tau-m^2\bigr]
\bigl[(1+a)(1-ax)-2\sigma/\zeta^2\bigr]}\Biggr],
\label{e:B11}
\end{split} 
\\
\begin{split} 
B^{\rm{real}}_{12}&(\tau,\sigma,x,\zeta^2)  = \frac{C_F\alpha_s}{2
\pi^2}\delta^+(\tau-\sigma+m^2-\lambda^2)
\\
& \times
\Biggl[\frac{(1+a)}{\bigl[\tau-m^2\bigr]\bigl[(1+a)(1-ax)-2\sigma/\zeta^2\bigr]}\Biggr],
\label{e:B12}
\end{split} 
\\
\begin{split} 
B^{\rm{real}}_{13}&(\tau,\sigma,x,\zeta^2) =  \frac{C_F\alpha_s}{2
\pi^2}\delta^+(\tau-\sigma+m^2-\lambda^2)
\\
& \times
\Biggl[\frac{-(1+a)}{\bigl[\tau-m^2\bigr]\bigl[(1+a)(1-ax)-2\sigma/\zeta^2\bigr]}\Biggr],
\label{e:B13}
\end{split} 
\\
B^{\rm{real}}_{14}&(\tau,\sigma,x,\zeta^2) =
0\, .
\label{e:B14}
\end{align} 
%
%
These results demonstrate that all terms in Eq.~\eqref{e:gThat_zeta}
contribute to generate a nonzero $\widehat{g}_T$ since
(i) the $B_i$ terms are nonzero,
(ii) the PCFs can depend explicitly on $x$, and
(iii) the boundary term at $\bm{k}_T^2 = \infty$ cannot be neglected.

\section{Quark target TMDs and PDFs at $x<1$} 
\label{a:model2}

We are now in a position to calculate the TMDs for a quark target
defined in Eqs.~(\ref{e:g1_AB_zeta})--(\ref{e:gT_AB_zeta}), their
$\bm{k}_T$-integrals appearing in the LIR of Eq.~\eqref{e:mod_LIR}, and
the function $\widehat{g}_T$ as defined in Eq.~\eqref{e:gThat_zeta}. 
Similar calculations have been performed in
\cite{Harindranath:1997qn,Kundu:2001pk,Ji:2004wu,Schlegel:2004rg,Schlegel:2004di}. 
Without entering into details, we note that the light-cone divergences
occurring for $\zeta \to \infty$ can be moved to $x=1$, introducing the
well-known ``plus'' distribution~\cite{Ji:2004wu,Bacchetta:2008xw}. 
If we restrict ourselves to the region $x<1$, the results are free of
light-cone divergences and do not depend on $\zeta$.  In this region
we can use either Eqs.~\eqref{e:g1_AB_zeta}--\eqref{e:gT_AB_zeta} or
\eqref{e:g1_AB}--\eqref{e:gT_AB}.  The resulting functions are then
given by
\begin{align} 
\begin{split} 
&g_{1L}(x<1,{\bm k}^2_T)  =  \frac{2 C_F
\alpha_s}{(2\pi)^2}\frac{1}{{\bm
k}^2_T+x\lambda^2+(1-x)^2m^2}
\\ & \times
\biggl[1-x-\frac{2(1-x)(1-x(1-x))m^2}{{\bm
k}^2_T+x\lambda^2+(1-x)^2m^2}+\frac{2x}{(1-x)_+}\biggr]
,\label{e:g1LQTM}
\end{split} 
\\
& g_{1T}(x<1,{\bm k}^2_T)  =  -\frac{2 C_F
\alpha_s}{(2\pi)^2}\frac{2x(1-x)m^2}{({\bm
k}^2_T+x\lambda^2+(1-x)^2m^2)^2}
,\label{e:g1TQTM}
\\
\begin{split} 
& g_{T}(x<1,{\bm k}^2_T)  =  \frac{2 C_F
\alpha_s}{(2\pi)^2}\frac{1}{{\bm
k}^2_T+x\lambda^2+(1-x)^2m^2}
\\
& \quad \times
\biggl[x-\frac{(1-x)^2(1+x)m^2}{{\bm
k}^2_T+x\lambda^2+(1-x)^2m^2}+\frac{1+x}{(1-x)_+}\biggr]
.\label{e:gTQTM}
\end{split} 
\end{align} 
When working with non-lightlike Wilson lines, it is not clear how to
obtain the collinear parton distribution functions upon integration over
the transverse momentum~\cite{Ji:2004wu}.
However, at the one-loop level these subtleties are relevant only at
$x=1$.
Since we restrict ourselves to the region $x<1$, we can safely compute
collinear PDFs through
$\bm{k}_T$-integration.
For simplicity we choose an upper boundary $Q$ for the
$\bm{k}_T$-integration, and shift quark mass effects into the finite
part by introducing an arbitrary infrared cutoff $\mu$ in order to
obtain agreement with the results of
Refs.~\cite{Harindranath:1997qn,Kundu:2001pk}.  The divergent parts of
the parton distributions, {\it i.e.}, the terms including the upper
cutoff $Q$, are given by
\begin{align}  
g_{1L}(x<1) &=
\frac{\alpha_s C_F}{2\pi}\,
\frac{1+x^2}{(1-x)_+}\ln\frac{Q^2}{\mu^2}
,\label{e:g1coll}
\\
g_T(x<1) &=
\frac{\alpha_s C_F}{2\pi}\,
\frac{1+2x-x^2}{(1-x)_+}\ln\frac{Q^2}{\mu^2}
,\label{e:gTcoll}
\\
g_{1T}^{(1)}(x<1) &= -\frac{\alpha_s C_F}{2\pi}
x(1-x)\ln\frac{Q^2}{\mu^2}
.\label{e:g1T1coll}
\end{align} 
These results have appeared earlier in
Refs.~\cite{Harindranath:1997qn,Kundu:2001pk,Ji:2004wu,Schlegel:2004rg,Schlegel:2004di},
but have been derived here for the first time starting from the PCFs.

For $\widehat{g}_T$ at $x<1$, using either Eq.~\eqref{e:gThat_zeta} or
Eq.~\eqref{e:gThat} we obtain
\begin{equation} 
\widehat{g}_T(x<1) =
\frac{\alpha_s C_F}{2\pi}\,\ln\frac{Q^2}{\mu^2}\, ,
\label{e:gThatcoll}
\end{equation} 
confirming the result in Eq.~\eqref{e:gThat_LIR}, which was not obtained
directly but rather using the LIR relation Eq.~\eqref{e:mod_LIR}.

\bibliographystyle{myrevtex}
\bibliography{mybiblio}

\end{document}